\documentclass[12pt]{article}

\usepackage{tikz}

\usepackage[algoruled,linesnumbered]{algorithm2e}
\usepackage{algorithmic}

\usepackage{amsmath,amssymb}
\usepackage{graphicx}
\usepackage{epstopdf}
\usepackage{caption}
\usepackage{color}
\usepackage{dcolumn}
\usepackage{bm}
\usepackage[numbers,super,comma,sort&compress]{natbib}
\usepackage{fancyhdr}
\usepackage{lmodern}

\definecolor{background-color}{gray}{0.98}

\title{Fifth Geometric--Arithmetic Index  VS   Atom--Bond Connectivity Index and Heat of Formation }
\author{
Matev\v z \v Crepnjak  \thanks{University of Maribor, Faculty of Natural Sciences and Mathematics, Koro\v ska cesta 160, 2000 Maribor, Slovenia; University of Maribor, Faculty of Chemistry and Chemical Engineering, Smetanova ulica 17, 2000 Maribor, Slovenia}
Petra \v Zigert Pleter\v sek \thanks{University of Maribor, Faculty of Natural Sciences and Mathematics, Koro\v ska cesta 160, 2000 Maribor, Slovenia; University of Maribor, Faculty of Chemistry and Chemical Engineering, Smetanova ulica 17, 2000 Maribor, Slovenia}}

\begin{document}

\maketitle

\begin{abstract}
The geometric--arithmetic indices are widely considered in  the chemical graph theory  in the last decade. The reason of introducing new indices is to gain  prediction of target properties of considered molecules that is better than  the prediction obtained by already known indices. In the case of the fifth geometric--arithmetic  index  hitherto no ability of prediction of some molecule property has been considered. 

In this paper we investigate correlations  between the fifth geometric--arithmetic index and some other degree--based topological indices on the family of octane isomers and  polyaromatic hydrocarbons. Since a very good correlation is established with the well known atom--bond connectivity index for polyaromatic hydrocarbons and then for the alkane series,   the relation 
between the  heat of formation and the  fifth geometric--arithmetic index is examined and a good correlation is confirmed in that case as well.

\end{abstract}

\bigskip
\noindent
\textcolor{red}{{\bf Keywords}: Fifth geometric--arithmetic index, atom--bond connectivity index, heat of formation, correlation.}

\clearpage


  \makeatletter
  \renewcommand\@biblabel[1]{#1.}
  \makeatother

\bibliographystyle{apsrev}

\renewcommand{\baselinestretch}{1.5}
\normalsize

\clearpage

\section*{\sffamily \Large INTRODUCTION} 

Among the {\em molecular descriptors} which provide information on molecular  constitution, 
{\em topological indices} have several assets such as: (1) easy calculation with very low computer time (cpu)
requirements; (2) diversity of possibilities to choose from   in order to match properties of the data set; (3) high
correlation ability with chemical  and physical properties or biological activities.
There are numerous of topological indices that have found some applications in theoretical chemistry,
especially in QSPR/QSAR research. Within all topological indices ones of the
most investigated are the descriptors based on the valences of atoms in molecules,  so-called degree--based topological indices.

Among the degree--based topological indices a class of {\em geometric--arithmetic} topological indices \cite{fu-gr-vu} may be  defined as
$$GA_{gen}(G)=\sum_{uv \in E(G)} \frac{2\sqrt{Q_u Q_v}}{Q_u+Q_v}\,,$$
where $Q_u$ is some quantity that  can be in a unique manner
associated with  vertex $u$ of  graph $G$. The first member of this class was
considered by Vuki\v cevi\' c and Furtula \cite{vu-fu} in year 2009 by setting $Q_u$ to be the vertex degree $d(u)$.
One year later  Fath-Tabar et al. \cite{fa-th}  introduced the second such index by setting $Q_u$ to be the number $n_u$, which is the number 
of vertices of G lying closer to  vertex $u$ than to  vertex $v$ for  edge $uv$ of a graph $G$.
The edge variant was studied by  Bo Zhou et al. \cite{zh-gu} in 2009 and lead to the third geometric--arithmetic index;   $Q_u$ 
being the number $m_u$ of edges of $G$ lying closer to  vertex $u$ than to  vertex $v$ for 
edge $uv$ of a graph $G$.
The fourth member of this class was considered by Ghorbani et al. \cite{gh-kh} in 2010 by setting $Q_u$
to be the eccentricity of vertex $u$ denoted by $\varepsilon_u$ and
finally the fifth geometric--arithmetic index was defined in 2011 by Graovac et al. \cite{gr_gh_ho} by letting $Q_u$ to be the sum of degrees
over all vertices incident with vertex  $u$.

Beside that the edge and the total versions of geometric--arithmetic indices were considered\cite{mah-1,mah-2} and recently Wilczek\cite{wi} defined nine new geometric--arithmetic indices.
Some mathematical properties  of first four geometric--arithmetic indices are obtained in \cite{das-1,das-2,das-3,das-4,das-5,das-6, ro-1,ro-2}.
It was  also shown  that first three geometric--arithmetic indices possess relatively good
descriptive as well as predictive capabilities with respect to some selected properties of
octanes and benzenoid hydrocarbons\cite{das-4, vu-fu}.

After the introduction of the fifth geometric--arithmetic index it has been  calculated for many different families of (molecular) graphs,  but there were no known correlation nor with physico-chemical  properties of molecules nor with any other topological indices. The aim of the present  paper is to fill this gap. The fifth geometric--arithmetic index is compared with some other distance-based topological indices and physico-chemical properties. As it is best correlated  with the atom--bond connectivity index, which is used for predicting the heat of formation 
of certain hydrocarbons, the connection between the fifth geometric--arithmetic index  and the heat of formation   is established.


\section*{\sffamily \Large THE FIFTH GEOMETRIC-ARITHMETIC INDEX AND SOME OTHER  DISTANCE-BASED TOPOLOGICAL INDICES}
\label{sec:indices}

A \textit{graph} $G$ is an ordered pair $G = (V, E)$ of a set $V$ of \textit{vertices} (also called nodes or points) together with a set $E$ of \textit{edges}, which are $2$-element subsets of $V$ (more information about some basic concepts in graph theory can be found in a book written by West\cite{west}). Having a molecule, if we represent atoms by vertices and bonds by edges, we obtain a \textit{molecular graph}.

The graphs considered in this paper are all finite and connected. The {\em degree} $d(u)$ of a vertex $u \in V(G)$ is the number of edges incident to  vertex $u$.

The fifth  geometric--arithmetic index is defined as 
 $$ GA_5(G)=\sum_{uv \in E(G)} \frac{2\sqrt{ S_u S_v}}{S_u+S_v}\,,$$
where $\displaystyle S_u=\sum_{uv \in E(G)} d(v)$.

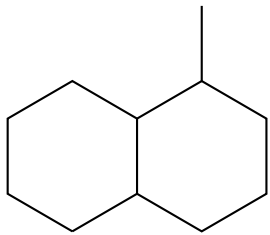
\begin{figure}
	\begin{center}
	\begin{tikzpicture}
		\draw[thick] (0,0) -- (0.86,0.5) -- (0.86,1.5) -- (0,2) -- (-0.86,1.5) -- (-0.86,0.5) -- (0,0);
		\draw[thick] (0.86,0.5) -- (1.72,0) -- (2.58,0.5) -- (2.58,1.5) -- (1.72, 2) -- (0.86,1.5);
		\draw[thick] (1.72, 2) -- (1.72, 3);
	\end{tikzpicture}
	\caption{Molecular graph $G$ of 1-methylnaphthalene.}
	\label{fig1}
	\end{center}
\end{figure}

As an example, we calculate the fifth geometric--arithmetic index for 1-methylnaphthalene (see Figure \ref{fig1}).

First we denote the molecular graph of 1-methylnaphthalene by $G$. Then
\begin{align*} 
\text{GA}_5 (G) & =  \frac{2\sqrt{4 \cdot 4}}{4+4} + 4\frac{2\sqrt{4 \cdot 5}}{4+5} + \frac{2\sqrt{5 \cdot 8}}{5+8} + 
									 \frac{2\sqrt{6 \cdot 8}}{6+8} + \frac{2\sqrt{5 \cdot 6}}{5+6} + 2\frac{2\sqrt{5 \cdot 7}}{5+7} + 
									 \frac{2\sqrt{7 \cdot 8}}{7+8} + \frac{2\sqrt{3 \cdot 6}}{3+6} \\
								& \approx 11,8465.
\end{align*}

The fifth geometric--arithmetic index was firstly computed for nanostar dendrimers \cite{gr_gh_ho}, followed by
 circumcoronene series\cite{fa3}, zig-zag polyhex nanotubes and nanostars\cite{fa2},
 $TURC4C8(S)$ nanotube\cite{fa-5}, armchair polyhex nanotubes\cite{fa1} and polyaromatic hydrocarbons\cite{fa-4}. Beside that this index was computed for naphtalenic nanosheet $[4n,2m]$\cite{am},  fan and wheel molecular graphs\cite{gao}, bridge and carbon nanocones\cite{ga_wa} and just recently for para-line graphs in convex polytopes\cite{fo}.

Probably the most studied degree--based topological indices  are the Zagreb indices which have been introduced  almost fifty years ago by Gutman and
Trinajsti\v c \cite{tr_gu}. The {\em first Zagreb index} is defined as 
$$ZM_1(G)=\sum_{v \in V(G) } d^2(v)$$
and the {\em second Zagreb index} equals
$$ZM_2(G)=\sum_{uv \in E(G) } d(u)d(v)\,.$$
Zagreb indices can also  be calculated by using the {\em valence vertex degree} $d^v(u)$ (i.e. the number of valence electrons of $u$ minus the number of hydrogen atoms attached to  $u$)  resulting  in the first and the second valence Zagreb indices
$$ZM_1^v(G)=\sum_{u \in V(G) } (d^v(u))^2 \quad \text{and} \quad ZM_2^v(G)=\sum_{uw \in E(G) }(d^v(u))^2 (d^v(w))^2\,.$$
Some other important degree--based topological indices are 
{\em Randi\'c connectivity index} \cite{randic}
$$\chi (G)=\sum_{uv \in E(G)} \frac{1}{\sqrt{d(u)d(v)}}\,,$$
the  {\em Pogliani index} \cite{pogliani}
$$Dz(G)=\sum_{u \in V(G) } d^Z(u)\,,$$
where $d^Z(u)$ is the  $Z$-delta number of $u$  and is defined as a quotient between the number of valence electrons and the principal quantum number of  vertex  $u$.
Further we have 
the {\em atom--bond connectivity index} \cite{es_to_ro_gu}
$$ABC(G)=\sum_{uv \in E(G)} \sqrt{\frac{d(u)+d(v)-2}{d(u)d(v)}}\,,$$
the {\em ramification index}  \cite{ar-pe}
$$Ram(G) = \sum_{v \in V(G), \, d(v)\geq 3} (d(v)-2)\,,$$
{\em Narumi simple index}  \cite{narumi} 
$$Snar(G) = \prod _{v \in V(G)} d(v)\,,$$
the {\em total structure connectivity index} \cite{ne_we_se}
$$Xt(G) =\prod_{v \in V(G)} \frac{1}{\sqrt{d(v)}}\,, $$
and  the {\em quadratic index} \cite{balaban},  also called normalized quadratic index
$$Q(G)=\frac{1}{2}\sum_{g} ((g^2-g)\,^gF+2)=3-2\vert V(G) \vert+\frac{M_1(G)}{2}\,,$$
where $g$ are the different vertex degree values and $^g F$ is the vertex degree count.

\section*{\sffamily \Large COMPUTATIONAL DETAILS}


In the present section we give an algorithm which we use to compute the fifth geometric--arithmetic index. The algorithm contains two special functions, i.e. {\tt CalculateDegree} and {\tt CalculateS}.

Let $G$ be a graph given by an adjacency matrix with vertices $1,2,\ldots,n$. The function {\tt CalculateDegree} computes degree for a given vertex. The function 
{\tt CalculateS} determines the value which is the sum of degrees of all neighbors for a given vertex. Note that this two special functions can be obtained easily and both functions have the time complexity $O(n^2)$. 
\medskip

\begin{small}

\begin{algorithm}[H]\label{alg:edini}
\SetKwInOut{Input}{Input}
\SetKwInOut{Output}{Output}
\DontPrintSemicolon

\Input{\! Graph $G$ with vertices $1,2,\ldots,n$.}
\Output{$\text{GA}_5(G)$}

	\SetKwFunction{CD}{CalculateDegree}
	\SetKwFunction{CS}{CalculateS}

  $X \leftarrow 0$\;
 
 	\For{\bf{each} $u \in V(G)$}{
		$d_u \leftarrow \CD(u)$\;
	
 	}
 	
 	\For{\bf{each} $u \in V(G)$}{
		$S_u \leftarrow \CS(u)$\;
 	}
 	
 	\For{\bf{each} $uv \in E(G)$}{
		$X \leftarrow X + \frac{2 \sqrt{S_u S_v}}{S_u + S_v}$\;	
 	}

	$\text{GA}_5(G) \leftarrow X$ \;

 \caption{The fifth geometric--arithmetic index.}
\end{algorithm} 
\end{small}
\medskip

The fifth geometric--arithmetic indices for the polyaromatic hydrocarbons molecules are collected in Tables \ref{tab:GA5+ABC} 
and \ref{tab:GA5+ABC2}; and for the alkane series in Table  \ref{tab:GA5+H}.

Since the data for the atom--bond connectivity index for the polyaromatic hydrocarbons is not available, we compute these values with a simple algorithm. The algorithm for calculating the atom--bond connectivity index is quite similar to the algorithm for calculating the fifth geometric--arithmetic index. For the completeness of the paper, we give it anyway.

As before, let $G$ be a graph given by an adjacency matrix with vertices $1,2,\ldots,n$ and  the function {\tt CalculateDegree} computes degree for a given vertex. Let us mention that Algorithm \ref{alg:ABC} has the time complexity $O(n^2)$.

\begin{small}

\begin{algorithm}[H]\label{alg:ABC}
\SetKwInOut{Input}{Input}\SetKwInOut{Output}{Output}
\DontPrintSemicolon

\Input{Graph $G$ with vertices $1,2,\ldots,n$.}
\Output{$\text{ABC}(G)$}

	\SetKwFunction{CD}{CalculateDegree}

  $X \leftarrow 0$\;
 
 	\For{\bf{each} $u \in V(G)$}{
		$d_u \leftarrow \CD(u)$\;	
 	}
 	
 	\For{\bf{each} $uv \in E(G)$}{
		$X \leftarrow X + \sqrt{\frac{d_u + d_v - 2}{d_u d_v}}$\;	
 	}

	$\text{ABC}(G) \leftarrow X$ \;

 \caption{The atom--bond connectivity index.}
\end{algorithm}
\end{small}
\medskip

The  atom--bond connectivity indices for the  polyaromatic hydrocarbons   are collected in Tables \ref{tab:GA5+ABC} and \ref{tab:GA5+ABC2}.



\section*{\sffamily \Large RESULTS AND DISCUSSION}


A benchmark data set for the octane isomers and the polyaromatic haydrocarbons  is available  at {\rm  www.moleculardescriptors.eu}.

For the  set of 18 octane isomers we  have compared the  fifth geometric--arithmetic   index with 16 physico-chemical properties and then further 
with the  degree--based topological indices - unfortunately no significant correlation could be established.

\begin{center}
\begin{figure}[h!]
\centering
\includegraphics[width=0.9\columnwidth,keepaspectratio=true]{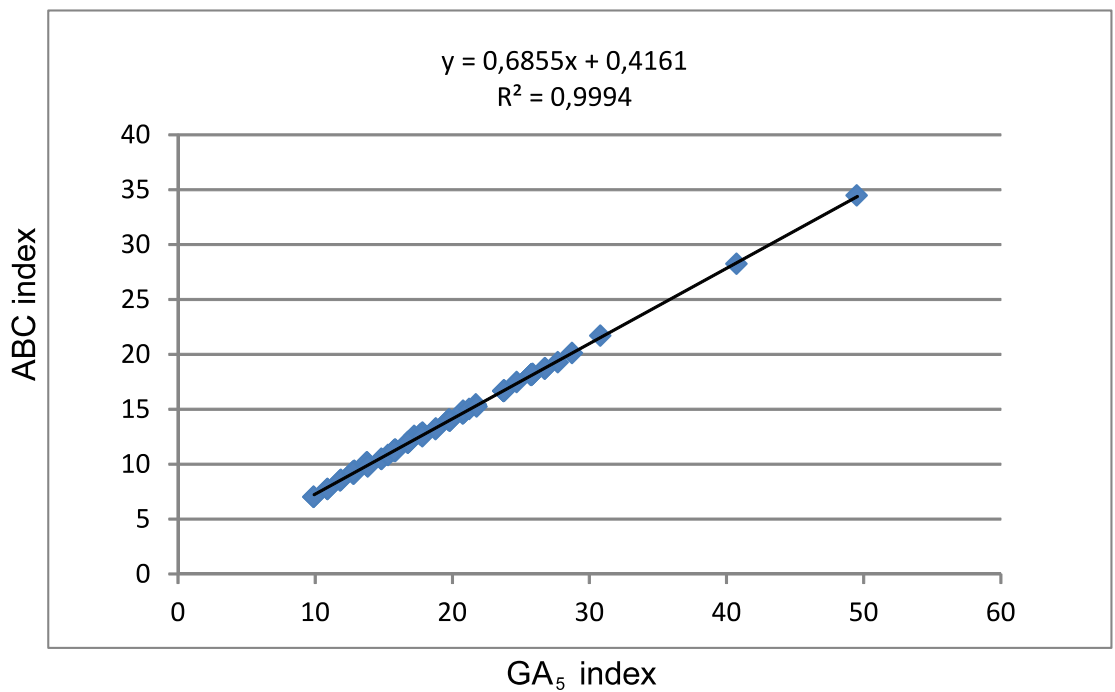}
\caption{\label{fig-ga5+ABC} The correlation between the fifth geometric--arithmetic index and the atom--bond connectivity index.}
\end{figure}
\end{center}

Next we  perform the regression analysis on  the set of 82 polyaromatic hydrocarbons. The benchmark data set enables the analysis of three physico-chemical properties (melting and boiling point, octanol-water  partition coefficient);  we could not establish any correlation between the fifth geometric--arithmetic index  and these properties.
Further  we consider degree--based indices and it turns out that the best correlation is the correlation between the fifth geometric--arithmetic index and the atom--bond connectivity index (see Figure \ref{fig-ga5+ABC}). More precisely, the regression statistics for these two indices is: multiple $R$ is $0,9997$,  $R^2$ is= $0,9994$; adjusted $R^2$ is 0,9994  and  standard error is 0,1621. Good correlation probably follows from the fact that the formulas for computing the fifth geometric--arithmetic index and the atom--bond connectivity index are quite similar. 

The second best correlation is the correlation between the fifth geometric--arithmetic index and the first valence Zagreb index. In this case, the regression statistics is as follows: multiple $R$ is 0,9984, $R^2$ is 0,9967; adjusted $R^2$ is 0,9967 and standard error is 0,3730. 
The correlation between the fifth geometric--arithmetic index and the Randi\'c connectivity index is also very good. More precisely, multiple $R$ is 0,9982; 	$R^2$ is 0,9964; adjusted $R^2$ is 0,9963 and standard error is 0,3928.
The correlation between the fifth geometric--arithmetic index is also very high with the following indices:  
Narumi simple index, Pogliani index, the first and the second Zagreb index, the second Zagreb valence index, and Harary index. One can see that the regression statistics for these indices is: multiple $R$ greater than 0,993, $R^2$ is greater than 0,987; adjusted $R^2$ is greater that 0,987 and standard error is at most 0,724.
The quadratic index, the ramification index, the total structure connectivity index also have a good correlation with the fifth geometric--arithmetic index. In this case the standard error is greater than 1, $R^2$ is between 0,94 and 0,98, multiple $R$ and adjusted $R^2$ are both  between 0,87 and 0,96. From this follows that correlation is still good, but in respect to other indices  it is relatively poor. In the case of  Schultz index  and Gutman index the correlation is weaker.
The regression statistics between the fifth geometric--arithmetic index and the  degree--based topological indices mentioned here are  all collected in Table \ref{tab:TheRegressionStatistics}.

As we can see the  fifth geometric--arithmetic index  correlates the best with the atom--bond connectivity index.
The atom--bond connectivity index was introduced in 1989 by Estrada et al.\cite{es_to_ro_gu} as a tool to describe the heat of formation
 $\Delta H$ of alkanes since it has shown  a good quantitative structure-property relationship (QSPR) model. We have therefore used the available data \cite{texas}  and checked if there is any correlation between the fifth geometric--arithmetic index and the heat of formation of certain polyaromatic hydrocarbons. The data   gathered in 
  Table \ref{tab:GA5+H} results in the  linear regression with  the multiple $R$ being 0,971 and $R^2$ equals 0,942, which is similar to the correlation obtained by
 by Gutman and Furtula \cite{gu-fu}, where the heat of formation of some polyaromatic hydrocarbons was compared with the first geometric--arithmetic index
and the established  correlation coefficient was 0,972.

Since the seminal paper \cite{es_to_ro_gu} on atom--bond connectivity index  models the heat of formation  $\Delta H$ of alkanes,  our next aim is to compare $\Delta H$ of alkanes 
with theirs  fifth geometric--arithmetic index. The data  in Table \ref{tab:GA5+H} is taken form   {\em www.webbook.nist.gov}  and results in a linear regression with the multiple $R=0,9999$ and $R^2=0,9998$.  The good correlation is due to a linear relation between  the fifth geometric--arithmetic index    and the atom--bond connectivity index  of the  alkane series. The molecular graph of an alkane  $C_nH_{2n}$  is a path  $P_n$ on $n$ vertices, so for  $n\geq 5$ a straightforward calculation yields in
$$\begin{array}{rcl}
GA_5(P_n) & = & 2(\frac{\sqrt{2\cdot 3}}{5}+\frac{\sqrt{3\cdot 4}}{7}+(n-5)\frac{\sqrt{4\cdot 4}}{8}+\frac{\sqrt{3\cdot 4}}{7}+
      \frac{\sqrt{2\cdot 3}}{5}) \\
&=&4\frac{\sqrt{6}}{5}+8\frac{\sqrt{3}}{7}+n-5\\
ABC(P_n)&=&(n-1)\frac{\sqrt{2}}{2}
  \end{array}$$
what  gives 
$$ABC(P_n)=0.7071\, GA_5(P_n)+0.0431\,.$$

\section*{\sffamily \Large CONCLUSIONS}

The fifth geometric--arithmetic index is an index in the family of the degree--based topological indices. The index is relatively new and  although its mathematical properties and closed formulae  for some families of chemical graphs were derived,  there is no proven relation between the fifth geometric--arithmetic index and physico-chemical properties or 
some other (degree--based) topological indices so far. 

In this paper we  consider  three types of molecules. In the case of octane isomers no significant results  could be shown. For the polyaromatic hydrocarbons and the alkane series
very good correlation and linear relation, respectively,  between the fifth geometric--arithmetic index and the atom--bond connectivity index is established. As a consequence the  fifth geometric--arithmetic index  is related with the heat of formation.  Our data set in that case consisted of 18 polyaromatic hydrocarbons and 19 members of the alkane series, what is not enough for a credible QSPR analysis and this is a problem which  could be  considered in the future by the chemical society, since the Algorithm \ref{alg:edini}  enables the calculation of the fifth geometric--arithmetic index.

\subsection*{\sffamily \large ACKNOWLEDGMENTS}

\noindent The authors Matev\v z \v Crepnjak and Petra \v Zigert Pleter\v sek acknowledge the financial support from the Slovenian Research Agency, research core funding No. P1-0403 and No. P1-0297, J1-9109, respectively.

\clearpage



\newpage

\begin{table}
\begin{small}
	\centering
		\begin{tabular}{|l|c|c|c|c|}
		\hline
		Index & Multiple R & R Square & Adjusted R Square & Standard Error \\		
		\hline
		Atom bond connectivity I	(ABC) &	0,9997 & 0,9994 & 0,9994 & 0,1621 \\
		\hline
		1.\ ZG valence I	(ZM1V) & 0,9984 &	0,9967 &	0,9967 &	0,3730 \\
		\hline
		Randi\'c connectivity I ($\chi$) & 0,9982 & 0,9964 & 0,9963 & 0,3928 \\
		\hline
		Narumi simple I (Snar) & 0,9982 & 0,9964 & 0,9964 & 0,3893 \\
		\hline
		Pogliani I (Dz) & 0,9974 &	0,9949 &	0,9948 &	0,4670 \\
		\hline
		First Zagreb I	(ZM1)	&	0,9975 &	0,9950 &	0,9949 & 0,4589 \\
		\hline
		2.\ ZG valence I (ZM2V) &	0,9964 & 0,9928 & 0,9927 & 0,5549 \\
		\hline
		Second Zagreb I	(ZM2)	& 0,9938 & 0,9877 & 0,9876 & 0,7232 \\
		\hline				
		Harary  I (Har) & 0,9972 &	0,9944	& 0,9944 &	0,4866 \\
		\hline
		Quadratic I (Q) & 0,9813 & 0,9630 & 0,9625 & 1,2564 \\
		\hline
	  Ramification I (Ram) & 0,9670 &	0,9352 &	0,9343	& 1,6628 \\
		\hline 
	  Total structure connectivity I (Xt) & 0,9348 & 0,8739 & 0,8724 & 2,3184 \\
		\hline \hline
		Schultz MTI (SMTI) & 0,9130	& 0,8335	& 0,8315	& 2,6641 \\
		\hline
		Gutman MTI (GMTI)	& 0,9066	& 0,8219	& 0,8196	& 2,7559 \\
		\hline		
		
		\end{tabular}
	\caption{The regression statistics between the fifth geometric--arithmetic index and other degree--based topological indices.}
	\label{tab:TheRegressionStatistics}
	\end{small}
\end{table}

\begin{table}
\begin{small}
	\centering
		\begin{tabular}{|l|c|c|l|c|c|}
		\hline
	Molecule	&	$GA_5$	&	$ABC$	&		Molecule	&	$GA_5$	&	$ABC$	\\
	\hline
1.	naphtalene	&	10,9193	&	7,73773	&	31.	4-5-methylenephenanthrene	&	17,854	&	12,5257	\\
2.	1-methylnaphthalene	&	11,8465	&	8,51379	&	32.	tetracene	&	20,8956	&	14,7279	\\
3.	2-methylnaphthalene	&	11,9068	&	8,55423	&	33.	benzo[a]anthracene	&	20,8547	&	14,6875	\\
4.	1-ethylnaphthalene	&	12,8067	&	9,11151	&	34.	chrysene	&	20,8069	&	14,647	\\
5.	2-ethylnaphthalene	&	12,8612	&	9,15195	&	35.	benzo[c]phenanthrene	&	20,8216	&	14,647	\\
6.	2-6-dimethylnaphthalene	&	12,8943	&	9,37073	&	36.	triphenylene	&	20,8009	&	14,6066	\\
7.	2-7-dimethylnaphthalene	&	12,8943	&	9,37073	&	37.	pyrene	&	18,832	&	13,2328	\\
8.	1-7-dimethylnaphthalene	&	12,8397	&	9,33029	&	38.	1-methylpyrene	&	19,7799	&	14,0089	\\
9.	1-5-dimethylnaphthalene	&	12,7781	&	9,28985	&	39.	2-methylpyrene	&	19,8264	&	14,0493	\\
10.	1-2-dimethylnaphthalene	&	12,7804	&	9,28985	&	40.	4-methylpyrene	&	19,7826	&	14,0089	\\
11.	1-3-7-trimethylnaphthalene	&	13,8272	&	10,1468	&	41.	2-7-dimethylpyrene	&	20,8208	&	14,8658	\\
12.	2-3-5-trimethylnaphthalene	&	13,7722	&	10,1063	&	42.	pentacene	&	25,8837	&	18,223	\\
13.	2-3-6-trimethylnaphthalene	&	13,8268	&	10,1468	&	43.	dibenzo[ai]anthracene	&	25,8429	&	18,1826	\\
14.	phenalene	&	14,8554	&	10,4853	&	44.	dibenzo[ah]anthracene	&	25,802	&	18,1421	\\
15.	1-phenylnaphthalene	&	17,8485	&	12,6066	&	45.	dibenzo[aj]anthracene	&	25,802	&	18,1421	\\
16.	2-phenylnaphthalene	&	17,8854	&	12,647	&	46.	benzo[b]chrysene	&	25,8007	&	18,1421	\\
17.	anthracene	&	15,9074	&	11,2328	&	47.	dibenzo[ac]anthracene	&	25,8005	&	18,1017	\\
18.	1-methylanthracene	&	16,8403	&	12,0089	&	48.	pycene	&	25,7529	&	18,1017	\\
19.	2-methylanthracene	&	16,8949	&	12,0493	&	49.	benzo[a]pyrene	&	23,8039	&	16,6875	\\
20.	2-7-dimethylanthracene	&	17,2661	&	12,567	&	50.	benzo[e]pyrene	&	23,7798	&	16,647	\\
21.	2-6-dimethylanthracene	&	17,8825	&	12,8658	&	51.	perylene	&	23,7639	&	16,647	\\
22.	2-3-dimethylanthracene	&	17,8275	&	12,8254	&	52.	coronene	&	27,7425	&	19,282	\\
23.	9-10-dimethylanthracene	&	17,6914	&	12,7041	&	53.	anthranthrene	&	26,8044	&	18,7279	\\
24.	phenanthrene	&	15,8609	&	11,1924	&	54.	benzo[ghi]perylene	&	26,7745	&	18,6875	\\
25.	1-methylphenanthrene	&	16,7925	&	11,9684	&	55.	dibenzo[ae]pyrene	&	28,7574	&	20,1017	\\
26.	2-methylphenanthrene	&	16,8484	&	12,0089	&	56.	1-methylchrysene	&	21,7594	&	15,4231	\\
27.	3-methylphenanthrene	&	16,8541	&	12,0089	&	57.	6-methylchrysene	&	21,2586	&	15,0006	\\
28.	4-methylphenanthrene	&	16,799	&	11,9684	&	58.	3-methylcholanthrene	&	24,7406	&	17,4635	\\
29.	9-methylphenanthrene	&	16,7945	&	11,9684	&	59.	indeno[1-2-3-cd]pyrene	&	26,7853	&	18,6875	\\
30.	3-6-dimethylphenanthrene	&	17,8473	&	12,8254	&	60.	pentaphene	&	25,8486	&	18,1826	\\
\hline
		\end{tabular}
	\caption{The fifth geometric--arithmetic index and atom--bond connectivity index for PAH's (part 1).}
	\label{tab:GA5+ABC}
	\end{small}
\end{table}

\begin{table}
\begin{small}
	\centering
		\begin{tabular}{|l|c|c|}
		\hline
	Molecule	&	$GA_5$	&	$ABC$	\\
	\hline
61.	hexaphene	&	30,8367	&	21,6777	\\
62.	indano	&	9,91929	&	7,03063	\\
63.	indene	&	9,91929	&	7,03063	\\
64.	azulene	&	10,9193	&	7,73773	\\
65.	acenaphthene	&	13,8678	&	9,77817	\\
66.	acenaphthylene	&	13,8678	&	9,77817	\\
67.	fluorene	&	14,8829	&	10,4853	\\
68.	1-methylfluorene	&	15,8202	&	11,2613	\\
69.	2-methylfluorene	&	15,8704	&	11,3018	\\
70.	3-methylfluorene	&	15,8761	&	11,3018	\\
71.	4-methylfluorene	&	15,3347	&	10,8388	\\
72.	9-methylfluorene	&	15,7793	&	11,2209	\\
73.	1-2-benzofluorene	&	19,8346	&	13,9399	\\
74.	fluoranthene	&	18,8156	&	13,1924	\\
75.	2-3-benzofluorene	&	19,8768	&	13,9804	\\
76.	3-4-benzofluorene	&	19,8436	&	13,9399	\\
77.	benzo[ghi]fluoranthene	&	21,8103	&	15,2328	\\
78.	benzo[k]fluoranthene	&	23,8152	&	16,6875	\\
79.	benzo[b]fluoranthene	&	23,7933	&	16,647	\\
80.	benzo[j]fluoranthene	&	23,7802	&	16,647	\\
81.	ovalene	&	40,7576	&	28,223	\\
82.	quaterryllene	&	49,5342	&	34,4657	\\
	\hline
		\end{tabular}
	\caption{The fifth geometric--arithmetic index and atom--bond connectivity index for PAH's (part 2).}
	\label{tab:GA5+ABC2}
	\end{small}
\end{table}

\begin{table}
\begin{small}
	\centering
		\begin{tabular}{|l|c|c||l|c|c|}
		\hline
Molecule	& heat of formation   & $GA_5$ &Molecule	& heat of formation   & $GA_5$ \\
                 &$\Delta H$   $[kJ/mol]$              &               &                  & $\Delta H$   $[kJ/mol]$                                   &\\
	\hline
naphtalene	 & 150,6	& 10,9193  & ethane & 20.4  & 1,000 \\
anthracene & 	227,7	& 15,9074  & propane & 25,02 & 2,000\\
phenanthrene	& 207,1	& 15,8609 & butane & 30,03 & 2,6200 \\
benzo[a]anthracene & 291	&20,8547 &  pentane & 35,08 & 3,9391  \\
chrysene	& 262,8	& 20,8069  & hexane & 39,96 & 4,9391 \\
benzo[c]phenanthrene	& 302,4	 & 20,8216 & heptane & 44,89 &  5,9391  \\
triphenylene	& 269,8	& 20,8009  & octane &  49,82 &  6,9391  \\
pyrene	& 225,7	& 18,832 & nonane & 54,66 &  7,9391 \\
pentacene	& 374	& 25,8837 & decane & 59,67 &  8,9391 \\
dibenzo[ah]anthracene	& 343	& 25,802 & undecane & 64,61 &  9,9391  \\
dibenzo[aj]anthracene	 & 343	& 25,802 & dodecane &  69,52 &  10,9391 \\
benzo[b]chrysene	& 346	& 25,8007  & tridecane & 74,45 &  11,9391 \\
dibenzo[ac]anthracene	& 345	& 25,8005  & tetradecane & 79,38	&	 12,9391 \\
pycene	& 334	& 25,7529 &pentadecane &84,30	&13,9391 \\
benzo[a]pyrene	& 301	& 23,8039& hexadecane & 	89,22	&	 14,9391 \\
benzo[e]pyrene& 	304	 & 23,7798  & heptadecane &94,2	17 & 15,9391 \\
perylene & 	324	& 23,7639& octadecane &99,08	&	 16,9391  \\
pentaphene	& 359	& 25,8486  & nonadecane & 	104,00	& 17,9391 \\
 && & icosane & 108,93	& 18,9391\\
	\hline
		\end{tabular}
	\caption{Heat of formation and  the  fifth geometric--arithmetic index and of some  PAH's  and the  alkane series. }
	\label{tab:GA5+H}
	\end{small}
\end{table}

\end{document}